\documentclass[]{aastex631}

\shorttitle{Origin and Composition of the Galactic Diffuse X-Ray Emission}

\shortauthors{Koyama \& Nobukawa}

\graphicspath{{./}{figures/}}

\begin{document}

\title{
Origin and Composition of the Galactic Diffuse X-Ray Emission Spectra by Unresolved X-Ray Sources}

\correspondingauthor{Masayoshi Nobukawa}
\email{nobukawa@cc.nara-edu.ac.jp}

\author{Katsuji Koyama}
\affiliation{
Department of Physics, Graduate School of Science, Kyoto University, \\
Kitashirakawa-oiwake-cho, Sakyo-ku, Kyoto 606-8502, Japan}

\author[0000-0003-1130-5363]{Masayoshi Nobukawa}
\affiliation{Faculty of Education, Nara University of Education, Takabatake-cho, Nara 630-8528, Japan}

\begin{abstract}

The Galactic diffuse X-ray emission (GDXE) can be spatially segmented into the Galactic center X-ray Emission (GCXE), the Galactic ridge X-ray emission (GRXE), and the Galactic bulge X-ray emission (GBXE).
The X-ray spectra of the GDXE are expressed by the assembly of compact X-ray sources, which are either the white dwarfs (WDs),  or  the X-ray active stars, consisting of binaries with late type stars.
The WDs  have either strong magnetic field (mCV), or weak magnetic field (non-mCV). The WDs and X-ray active stars are collectively called as compact X-ray stars. 
However, spectral fittings by the assembly of all compact X-ray stars for the GCXE, GRXE, and GBXE are rejected leaving 
significant excess near the energies of K$\alpha$, He$\alpha$,  Ly$\alpha$ lines.
These excesses are found in the collisional ionization equilibrium (CIE) plasma.
Thus the spectra of the GRXE and GBXE are improved by adding CIE-SNRs.
However the GCXE spectrum is still unacceptable with significant data excess due to
the radiative recombination emission  (RP-plasma). 
Then the GCXE fit is significantly improved by adding aged RP-SNRs.
The aged RP-SNRs would be made by a past big flare of Sgr~A$^*$ emitting either hard X-rays or low-energy cosmic-rays.
The big flares may excite Fe and Ni atoms in cold diffuse gas (CM), and emit fluorescent X-ray lines.
The CIE-SNRs, RP-SNRs and CM are called as diffuse X-ray sources.
This paper presents the spectral fits by the assembly of all the compact and diffuse X-ray sources together with high quality spectra 
and combined fit among all GDXE of GCXE, GRXE, and GBXE. This provides the 
first scenario to quantitatively and comprehensively predict the origin of the GDXE spectra.

\end{abstract}

\keywords{Galactic center(565) --- Diffuse x-ray background(384) --- Interstellar thermal emission(857) --- Supernova remnants(1667)}

\section{Introduction} \label{sec:intro}

X-ray view of our universe is revealed by \cite{Giacconi79} with the rocket born X-ray detectors. They discovered bright X-ray sources either point like or extended objects in our milky way Galaxy. In priority of this discovery, our view of the Universe is limited in the visible and infrared, and radio band.  
Well known X-ray sources has been super massive blackhole (SMBH) called as active galactic nuclei (AGN) and quasi-stellar object (QSO), and clusters of galaxies (CG).  The assembly of AGNs, OSQs, and CG are comprised as the cosmic X-ray background (CXB).  

In our Galaxy near the sun (within $\sim5$~kpc from us), notable point like X-ray sources are white dwarf binaries (WD) called cataclysmic variables (CV) with strong or moderate magnetic field (called as magnetic CV: mCV, and non-magnetic CV: non-mCV), and aged supernova remnant (aged-SNR).  These X-ray sources become invisible by the absorption by the high column density of interstellar medium (ISM) at the long distance near the Galactic center region of  $\gtrsim8.5$~kpc. 
However, \cite{Wo82} and \cite{Wo83} noted diffuse X-ray emission in the inner Galactic plane excluding the Galactic Center region, and was named as the Galactic Ridge X-ray emission (GRXE).  
Unlike the CXB, the origin of the GRXE has been long unresolved question.  What are the composition component among the known Galactic X-ray sources such as X-ray active stars (compact sources) or diffuse objects like a supernova remnant (SNR) ?    

 \citet{Mu04, Uc13, Ko18} found that the GDXE spectra can be classified into the spectra of GCXE, GRXE, and that of GBXE.
Fundamentally, the origin of the GDXE have been attributed to the assembly of unresolved X-ray sources \citep{Re06, Ho12}.
Based on the similarity of the spectra of GDXE in the 2--10~keV band, the magnetic cataclysmic variables (mCV) and non-magnetic CVs (non-mCV) have been regarded as the candidate sources.
The XABs are close binary systems consisting with low-mass stars and  enhanced coronal active stars. These are  also speculated as a good candidate of the GDXE compositions. The non-mCV, mCV, and XABs  are collectively referred to as X-ray active stars (XAS).
Then a crucial question related to the GDXE origin is ``how well can the GDXE spectra be reproduced by the assembly of observed real XASs spectra ?''. For a long time, no combination of the XAS spectra has succeeded to reproduce the GDXE spectra.  
For the GCXE spectrum, \citet{Mu04} collected point-source spectra from the data of the {\it Chandra} satellite in the $9'$-circular region surrounding the Galactic center black hole (Sgr~A$^*$), suggesting that the main components are non-mCVs and mCVs, with non-negligible contribution of the XABs. 
\citet{Uc13} extensively analyzed the {\it Suzaku} data of K-shell lines of S--Fe near the GCXE region, determining that the GCXE spectrum exhibits an aged SNR with recombining plasma.

The first quantitative analyses of the GCXE spectra were reported by \citet{Nob21} (hereafter, NK21). 
They extensively analyzed the {\it Suzaku} data of K-shell lines of S--Fe near the GCXE region, determining that the GCXE spectrum exhibits an aged-SNR of recombining plasma (RP).
In this paper, we extend NK21 to the GRXE and GBXE spectra in the wider X-ray band of 2--10~keV, focusing on the K-shell line structure of heavy elements such as S--Fe.
Since the spectral quality of the XAS has been limited in sample numbers near the solar vicinity (e.g., \cite{No16, Ko18} and their references), this study acquired extensive data set from the entire observation period (2005-2013) 
with the {\it Suzaku} satellite. 
The spectral analysis was carefully conducted with variable time linearity and resolution in the spectral energy along with the K-shell line energy of S--Fe. 
By including these processes, this paper presents the best quantitative analyses and results for the GDXE spectra which is not tried in any previous GDXE papers.  

\section{Observation and Data Reduction} \label{sec:obs} 
  
To investigate the origin of GDXE, we utilized the {\it Suzaku} archive data from the X-ray imaging spectrometer (XIS: \citealt{Ko07}) on the focal plane of the thin-foil X-ray telescopes (XRT: \citealt{Se07}). 
The {\tt Suzaku} data used in this study for XAS and GDXE (GCXE, GRXE, and GBXE) are listed in Tables~\ref{tab:obsXAS} and \ref{tab:obsGDXE}, respectively (see, Appendix).
XAS sources are referred to \cite{Xu16, No16}.
Moreover, we used cleaned event data after nominal screening methods supplied by the {\it Suzaku} team.
The analysis tools are taken from {\tt HEAsoft}~6.29 and relevant calibration databases.
The non-X-ray background data generated by {\tt xisnxbgen} \citep{Ta08} are subtracted from the raw spectra. 
The redistribution function files and auxiliary response files are generated by {\tt xisrmfgen} and {\tt xissimarfgen} \citep{Ishi07}, respectively.  In the spectral fitting, the software  package of {\tt XSPEC} (version 12.12.0) is used.

\section {Analysis and Results} \label{sec:ana-xas} 

The primary X-ray sources to contribute the origin of GDXE are XABs, non-mCVs, and mCVs (XAS).  
We  therefore make  the X-ray spectra of all the XASs from the Suzaku archives.  The list of the XAS data are given in Table~\ref{tab:obsXAS} (Appendix).
Then these spectra summed according each category (XAB, non-mCV, mCV) and are fitted  with a simple  thermal plasma model of {\tt Apec}. 

In the fitting, the electron temperature $kT_{\rm e}$ and abundance $z$ are free parameter. 
In all the XASs, the energy bands are 2--9~keV, which is wider than the previous pioneering study \citep{No16}  of 5--10~keV band. 
Using the best-fit spectrum of each XAB, non-mCV, and mCV (XAS), we obtained the luminosity-weighted average spectra, and fitted it with the model of {\tt Apec}. 

Since the spectra of XASs include fluorescent lines at 6.4 keV  from the stellar surface,  
we added a Gaussian line at 6.4~keV with a free equivalent width of $EW_{6.4}$.
The lower limit of $kT_{\rm e}$, and $z$ were set to be $\sim 1$~keV (refer to Figure~3 in \citealt{No16}), and  1-solar, respectively. These initial parameters are taken
 because the sample of the XASs are $\sim 1$~solar mass stars located  at $\sim 8$~kpc from the sun.

With these constraints, the spectra of XASs are fitted. 
The best-fit parameters are summarized in Table~\ref{tab:XASfit}. Hereafter quoted statistical errors are all 1$\sigma$ level estimated by {\tt error} command in {\tt XSPEC}. The best-fit figures and parameters are given in Figure 1 and Table 1. 

The best-fit parameters  are within the variation of each XASs best-fit. 
The best-fit abundance ($z$) are $\sim 0.3$, $\sim 0.9$, and $\sim 0.4$~solar for the XAB, non-mCV, and mCV, respectively (refer to Table~\ref{tab:XASfit}). These sub-solar abundances in XASs are unsolved mystery, but is beyond the scope of this paper. Important fact is that $z$ of non-mCV (0.87 solar) is far larger than that of mCV (0.42 solar), confirming the early suggestions  by \cite{Mu93}.  

The key point  of this paper is to check a complicated model with many free parameters by $\chi^2$ analysis. 
Then the logical process of the analysis starts from the simplest  model consisting of XAB, non-mCV and mCV (model A). The results of this model are given in the following subsection 3.1, which is previously supported by many authors. The next process is to add the CIE-SNR (model B). The results of the model B are given in subsection 3.2. This model improved the spectral fitting drastically, but still marginally rejected.
Then the model B is more improved by further addition of RP-SNR and CM (model C; subsection 3.3). The results of model C are largely improved. Namely the best-fit reduced $\chi^2$ value for GRXE and GBXE becomes well acceptable (1.50 and 1.59) but that for GCXE is at a still marginal level (1.80). 

Evolutions of the best-fit key spectral parameters of $kT_{\rm e}$ and $z$ in XASs (XAB, non-mCV and mCV) along the models A, B and C are obtained. The $kT_{\rm e}$ values of the XASs show large variations from models A, B and C in all XASs. However important fact is those of $z$ show no large variations. This fact indicates the line structures of Ni and Fe and S and Ca are constant in the models~A, B, and C.

The results in Table 1 are used as initial parameter values in the model A, B, and C fits.
These are not the final best-fit values but only initial parameter values. Therefore the smaller errors in Tables~2--4 than those of Table 1 are not the surprising issues.

\begin{table}[t!] 
\caption{Best-fit parameters of XASs (XAB, non-mCV, and mCV).}
\label{tab:XASfit}
\smallskip
\footnotesize
\begin{center}	
\begin{tabular}{lcccc}
\hline
			&	& XAB 			& non-mCV		& mCV \\
\hline		
{\tt Absorption} ($N_{\rm H}$) & $10^{22}$~cm$^{-2}$		&$<0.0016$                	&$0.118\pm0.026$	&$4.116\pm0.036$ \\
$kT_{\rm e}$	& keV	&$3.3\pm{0.1}$		& $7.4\pm{0.2}$	&$11.0\pm0.5$  \\
abundance ($z$) 	& solar	& $0.31\pm{0.01} $		& $0.87\pm{0.04}$	&$0.42\pm0.05$ \\
$EW_{6.4}$		& eV	& 43				& 100			&160 \\
\hline
$\chi ^2$/d.o.f.	&	& 1024/412(2.49)		& 451/189(2.39)			& 4573/261(17.5) \\
\hline
\end{tabular}
\end{center}
\tablenotetext{a}{Spectral parameters of XASs were luminosity-weighted average.
The abundances are all sub-solar (e.g., \citet{Pa12, By10, Ishi99} for XAB, non-mCV, and mCV, respectively). Therefore, the relative abundances in the table of \citet{An89} are stated. }
\end{table}

\subsection{GDXE spectral fit by XASs (model A)} \label{sec:modelA} 

The GDXE spectra are derived from the sum of several pointing observations (Table~\ref{tab:obsGDXE} in Appendix), with a total accumulation time of $\sim 1300$~ks, $\sim 800$~ks, and $\sim3000$~ks for the GCXE, GRXE, and GBXE, respectively. 
The energy resolutions of XIS in the early observation period are not degraded by particle background (cosmic-rays), whereas those of the late observation period are significantly degraded.
These time-dependent variations in the energy resolution and linearity are due to  random selection of data point
in the long exposure observations spanning from 2005 to 2013. 

The largest value of linearity variation is $\sim 15$~eV in the energy band of the K-shell lines of Fe and Ni (6--7~keV).
To compensate these line-broadening and nonlinearity effects, we apply {\tt Gsmooth} and {\tt gain} to the model fitting. 
The GDXE spectra exhibited line-like residuals at $\sim 3.2$ keV, which are possibly caused by the Au M-edge (from the X-ray mirror surface) \citep{Ku07}. 
To represent this line-like structure, we use a Gaussian line at $\sim 3.2$~keV with $\sigma=0.1$~keV according to \citet{Ku07}.

The conventional scenario in the early research for the origin of the GDXE is that the spectra are assembly of each XAS spectra.
However, this scenario has not been quantitatively evaluated (e.g., \citealt{Mu93, Wa14, Yu12, Xu16}). 
Thus, for the quantitative analysis, we fit the GDXE spectra following the same analysis procedure as NK21 but in more detail in wider X-ray band (referred as model~A). 
In this subsection of 3.1 and in the other subsections of 3.2 and 3.3, 
each XAS components in the GCXE, GRXE, and GBXE are simultaneously fitted.  

In the model~A, the electron temperature ($kT_{\rm e}$), abundances ($z$), and absorption ($N_{\rm H}$) of each XAS are free parameters.
In the XABs (Table 1), the best-fit  $kT_{\rm e}$ of $\sim3.3$~keV (Table~\ref{tab:XASfit}) is higher than the normal XABs (e.g., \citealt{Pa12}).                                             
This is because the important fitting energy band are within 5--10~keV (in \citealt{No16}), which is higher than the typical value of XABs of $kT_{\rm e}=1$--3~keV. 
Therefore, the initial value of $kT_{\rm e}$ of XABs is assumed  to be $\sim 1$~keV.

\begin{figure}[hbt]
\figurenum{1}
\epsscale{0.33}
\plotone{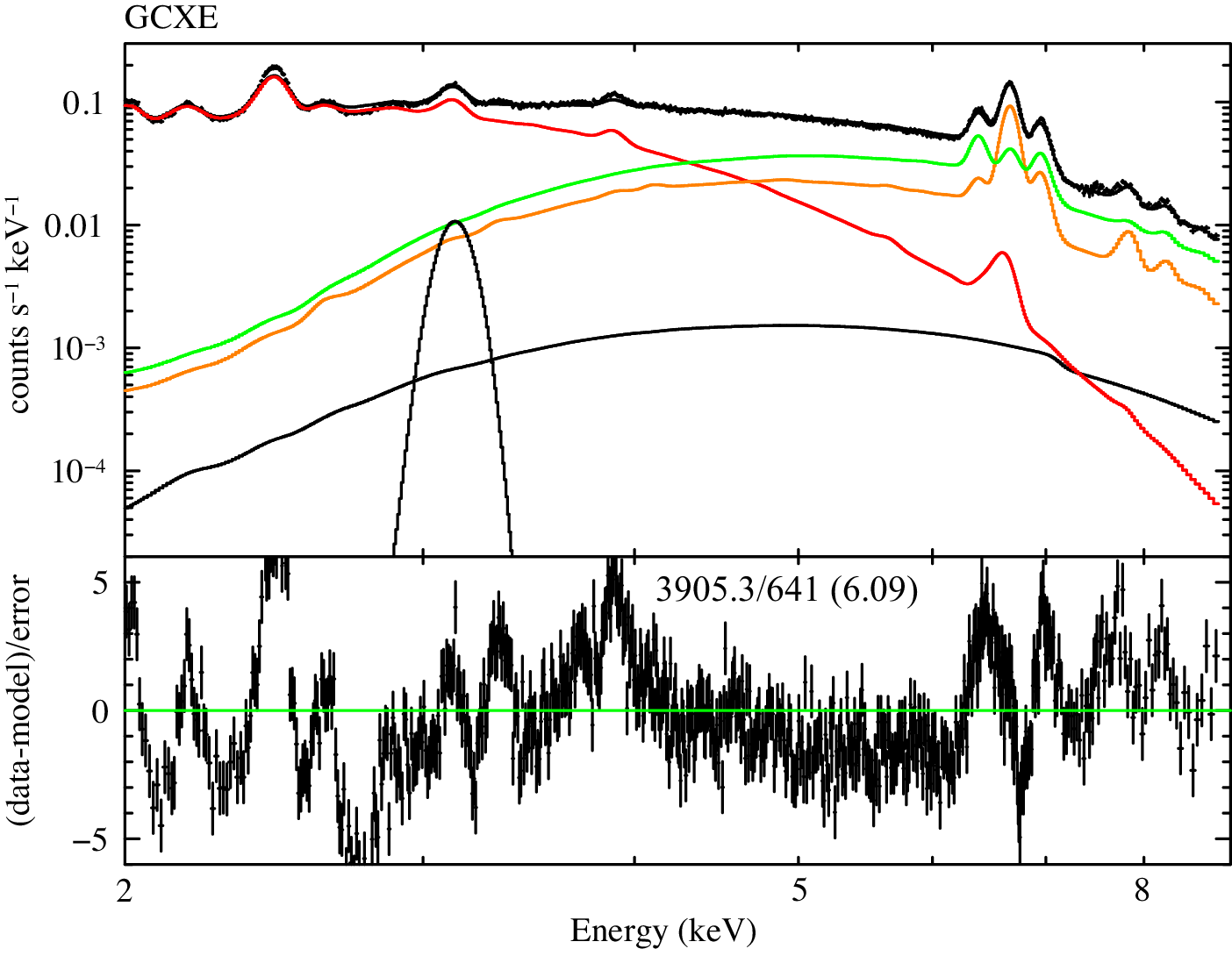}
\plotone{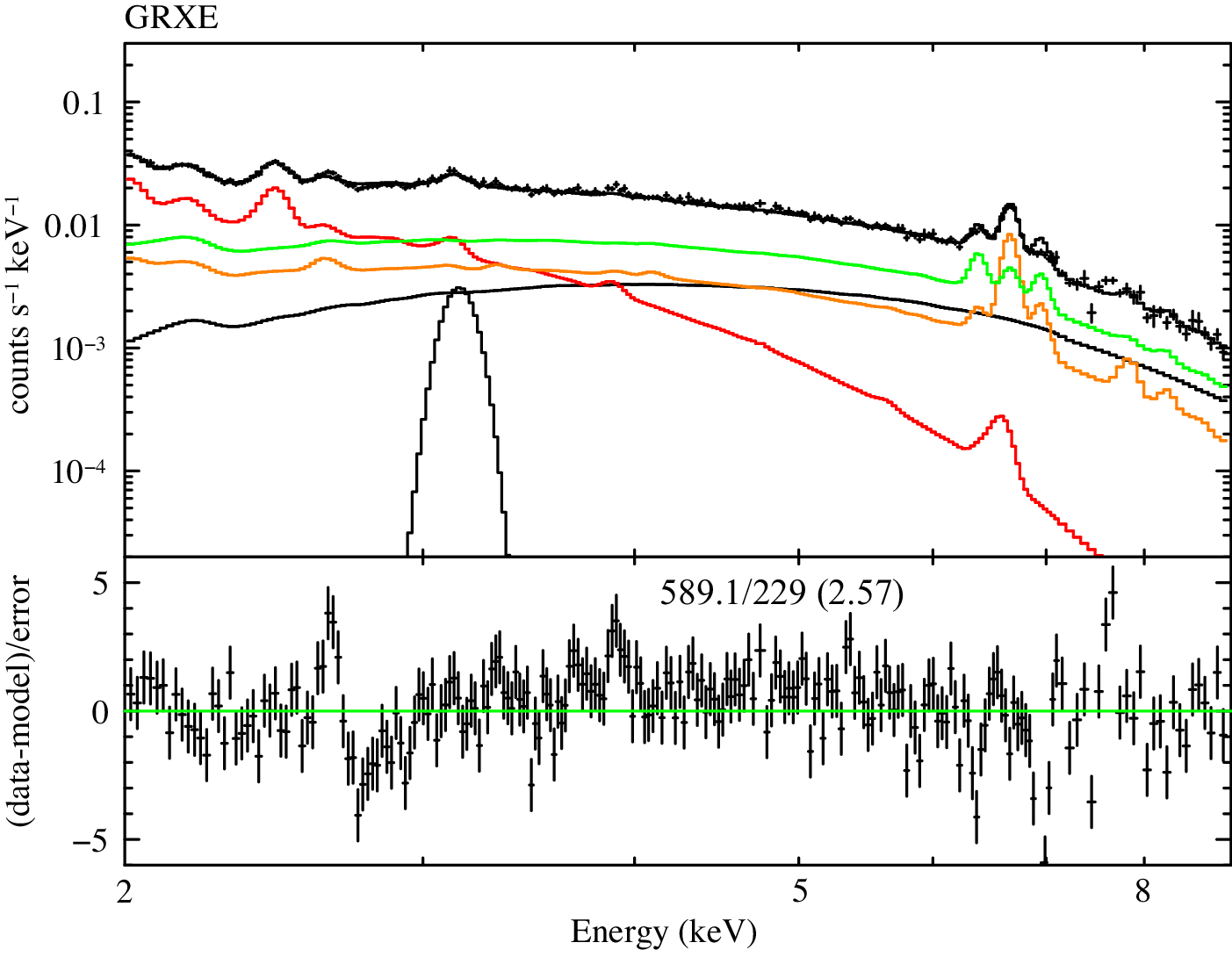}
\plotone{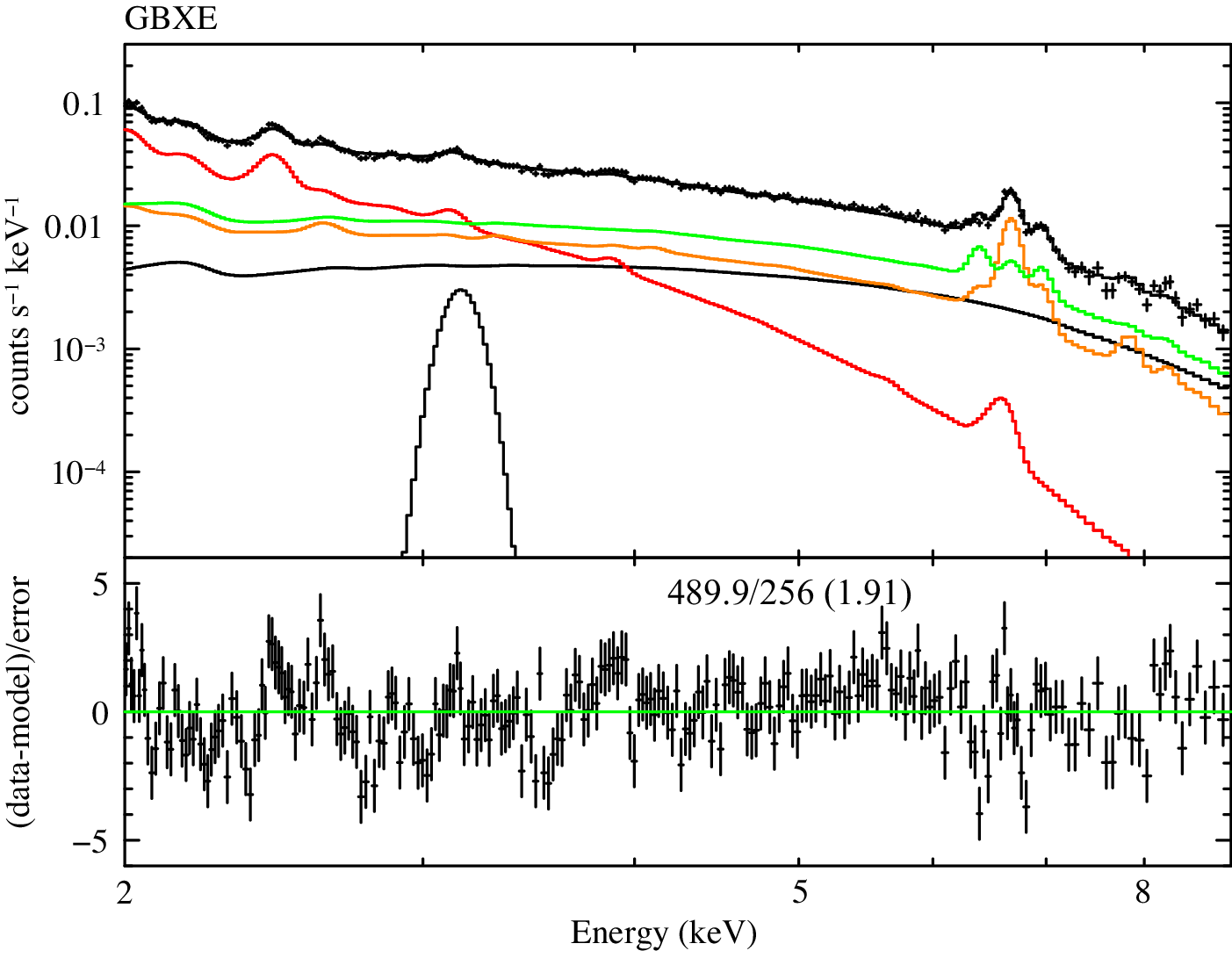}
\caption{
Top: GDXE (GCXE, GRXE, GBXE) spectra with the model~A, the combination of XAB (red), non-mCV (orange), and mCV (green).
 Black power-law model exhibits the CXB.
Bottom: Residuals between spectrum and model, including a $\chi ^2$/d.o.f. (reduced~$\chi^2$) value.
}
\label{fig:modelA}
\end{figure}

With these constraints, the spectra of GCXE, GRXE, and GBXE are simultaneously fitted. 
The best-fit model A figures and parameters of the GCXE, GRXE, and GBXE are given in Figure~\ref{fig:modelA} and Table~\ref{tab:modelA}, respectively.
The model A fit is statistically rejected  with $\chi^2$/d.o.f. (reduced $\chi^2$) of GCXE, GRXE, and GBXE of 6.09, 2.57, and 1.91, respectively. This  supports the early suggestions by many authors.  
In particular, the large data residuals in GCXE are found at the energies of iron K-shell lines at 6.4 keV, 6.68 keV, and 6.97 keV as well as the energies of S K-shell lines of 2.46~keV and 2.62~keV.  Further more complex residuals near at 3--4.5 keV are also found. 

The best-fit $\chi^2$/d.o.f. of mCV in Table~1 is  obtained by the spectrum fit, where the fit is made in the energy range of 2--9~keV.
The initial parameters of this fit are those of the results in NK21, in which 
the fitting energy band is 5--10 keV.
The large reduced $\chi^2$ (6.09) of the model A fit for GCXE (left figure in Figure~1) is due to the large data excess in the K-shell lines of Fe and Ni (above $\sim 6$~keV)
 and those of lighter elements S--Ca (in the 2--4.5~keV band) (see also NK21).
The excess of mCV in the GCXE, GRXE and GBXE is compensated by the other spectra of non-mCV and XAB in the simultaneous fit in the 2--9~keV range.
Thus, the error values in Table~2 are smaller than those of Table~1.

\begin{table*}[tbh] 
\caption{Best-fit parameters of GCXE, GRXE and GBXE in model~A.}
\label{tab:modelA}
\smallskip
\footnotesize
\begin{center}	
\begin{tabular}{llccccc}
\hline
&	&				&					& GC			& GR			& GB	\\
\hline
			&{\tt Absorption} & ${N_{\rm H}} $ 	& $10^{22}$~cm$^{-2}$	& $6.56\pm 0.01$	& $2.54\pm 0.01$	& $1.12\pm0.01$	\\
\hline
{\bf XAB}	&{\tt Apec}	& $kT$	& keV				& $1.0\pm0.1$		& =GC			& =GC	\\
			&		& $Z^a$		& solar				& $0.32\pm0.01$		& =GC			& =GC	\\
			&		& $EM^b$	& $10^{-2}$			& $138\pm 3$		& $1.3\pm 0.1$	& $2.6\pm 0.1$	\\
			& Fe~K$\alpha$		& $EW_{6.4}$		& eV				& $43$			& =GC		& =GC	\\
\hline
{\bf non-mCV}	&{\tt Apec}		& $kT$				& keV				& $5.5\pm0.1$	& =GC			& =GC	\\
			&		& $Z^a$		& solar				& $1.10\pm0.01$		& =GC			& =GC	\\
			& 		& $EM^b$	& $10^{-2}$			& $9.0\pm 0.8$		& $0.13\pm 0.01$ &$0.28\pm 0.03$	\\
			& Fe~K$\alpha$	  	& $EW_{6.4}$		& eV				& $100$			& 	=GC			& =GC	\\
\hline
{\bf mCV}		&{\tt Apec}		& $kT$		& keV	& $11.2\pm0.1$		& 	=GC			& =GC	\\
			& 		& $Z^a$		& solar				& $0.41\pm0.01$		& 	=GC			& =GC	\\
			& 		& $EM^b$	& $10^{-2}$			& $12.4\pm 0.5$		& 	$0.22\pm 0.01$	& $0.36\pm 0.02$	\\
			& Fe~K$\alpha$		& $EW_{6.4}$		& eV				& $160$			& 	=GC			& =GC	\\
\hline
\multicolumn{4}{c}{$\chi^2$/d.o.f.}			& 3905/641		& 589.1/229		& 489.9/256 \\
\hline
\end{tabular}
\end{center}
\tablenotetext{a}{Relative to abundances in table of \citealt{An89}.}
\tablenotetext{b}{Emission measure of plasma model, $\frac{10^{-14}}{4 \pi D^2} n_{\rm e} n_{\rm H} V$, where $D$, $n_{\rm e}$, $n_{\rm H}$, and V denote the distance, electron density, hydrogen density, and volume, respectively.}
\end{table*}

\subsection{GDXE spectra fitted with XASs and aged CIE-SNRs (model B)} \label{sec:modelB} 

As are already shown in Figure 1, the high-quality spectral analysis for GCXE by the the model A (subsection~\ref{sec:modelA})  exhibited large reduced $\chi^2$ with unacceptable level of large residuals of
 K-sell lines of S and Fe (K$\alpha$, He$\alpha$ and Ly$\alpha$), and complex structures near at 3--4 keV.
The former K-shell line structures are likely from Fe in SNRs and later complex would be  also due to K-shell lines of 
Ar--Ca. These structures are likely from aged CIE-SNRs, which has not been observed before. 

\citet{Ko86} proposed that the assembly of unresolved aged CIE-SNRs contribute to these residuals.
Accordingly, the aged SNR in collisional ionization equilibrium (aged CIE-SNR) is added to the model~A.
This is called as model~B. 
In the model~B fitting, the initial values of electron temperature ($kT_{\rm e}$) and abundance ($z$) 
are set to be free parameters of  nearly the same values as those in the model~A fitting.

The aged-SNRs must escape from the detection as known SNRs. Therefore, in order to be the aged CIE-SNRs are 
newly resolved-SNRs, these must be largely extended and should exhibit low surface brightness with low temperature of $\lesssim1 $~keV. 
In fact, \citet{Ko86} numerically simulated for the spectra of aged-SNRs, and concluded that the ages are old enough 
as $ > 10^{4-5}$~years, and their temperature should be sufficiently low as $kT_{\rm e} \lesssim 1$ keV.
 All these requirements are satisfied in the observed results of the aged-SNRs near GC \citep{Po15}. 
Furthermore, the high-quality spectra of the aged CIE-SNRs in the GC Region with {\it Suzaku} exhibits over 1~solar abundances such that $z$ of Si--Ar were $\sim 1.3$~solar (G359.12$-$0.05; \citealt{Na10}) and $\sim 1.7$~solar 
(G359.41$-$0.12; \citealt{Ts09}). 

Thus, the initial value of $z$ in the aged CIE-SNRs is set to be slightly over 1-solar unlike the sub-solar abundances of XASs in Table~1.
The dissimilar constraint on $z$ between XASs and aged CIE-SNRs can be attributed to their dissimilar origins. 
The best-fit parameters of the GCXE, GRXE and GBXE are given in Table~\ref{tab:modelB}.
The simultaneous fit of the GCXE, GRXE, and GBXE are improved largely from model A fit with reduced $\chi^2$ 
of 2.62, 2.16 and 1.49, respectively (see Figures 2 and 3).  However the model B is still 
marginally unacceptable. Moreover, the best-fit $z$ for non-mCV
of 1.10 (Table 2) is significantly larger than that of initial value of 0.87 (Table 1).

\begin{table*} 
\caption{Best-fit parameters in model B.}
\label{tab:modelB}
\smallskip
\footnotesize
\begin{center}	
\begin{tabular}{llccccc}
\hline
&	&				&					& GC			& GR			& GB	\\
\hline
			& {\tt Absorption}	& ${N_{\rm H}} $ 	& $10^{22}$~cm$^{-2}$	& $6.20\pm 0.01$	& $2.31\pm 0.01$	& $0.83\pm0.01$	\\
\hline
{\bf XAB}		&{\tt Apec}	& $kT$		& keV				& $1.5\pm0.1$		& =GC			& =GC	\\
			&		& $Z^a$		& solar				& $0.32\pm0.01$	& 	=GC			& =GC	\\
			&		& $EM^b$	& $10^{-2}$				& $17.2\pm 6.3$	& $0.61\pm 0.11$	& $0.79\pm 0.17$	\\
			& Fe~K$\alpha$	& $EW_{6.4}$	& eV				& $43$			& 	=GC			& =GC	\\
\hline
{\bf non-mCV}	&{\tt Apec}	& $kT$		& keV				& $7.2\pm0.1$	 	& =GC			& =GC	\\
			&		& $Z^a$		& solar				& $1.10\pm0.01$	& =GC			& =GC	\\
			& 		& $EM^b$	& $10^{-2}$				& $11.4\pm 2.5$	& $0.14\pm 0.04$	& $0.34\pm 0.08$	\\
			& Fe~K$\alpha$	& $EW_{6.4}$	& eV				& $100$			& 	=GC			& =GC	\\
\hline
{\bf mCV}		&{\tt Apec}	& $kT$		& keV				& $20.0\pm0.1$	& 	=GC			& =GC	\\
			& 		& $Z^a$		& solar				& $0.27\pm0.01$	& =GC			& =GC	\\
			& 		& $EM^b$	& $10^{-2}$				& $3.9\pm 2.4$		& $0.15\pm 0.02$	& $0.22\pm 0.06$	\\
			& Fe~K$\alpha$	& $EW_{6.4}$	& eV				& $160$			& =GC			& =GC	\\
\hline
{\bf CIE-SNR} 	&{\tt vApec}	& $kT$		& keV			& $0.88\pm0.01$	& =GC			& =GC	\\
			&		& $Z^a$		& solar				& $1.0\pm0.1$	 	& =GC			& =GC	\\
			&		& $EM^b$	& $10^{-2}$				& $59.2\pm 8.6$	& $0.23\pm 0.07$	& $0.67\pm 0.14$	\\
\hline			
\multicolumn{4}{c}{$\chi^2$/d.o.f.}			& 1677.9/641		& 495.5/229		& 380.7/256 \\
\hline
\end{tabular}
\end{center}
\tablenotetext{a}{Relative to abundances in table of \citealt{An89}.}
\tablenotetext{b}{Emission measure of plasma model, $\frac{10^{-14}}{4 \pi D^2} n_{\rm e} n_{\rm H} V$, where $D$, $n_{\rm e}$, $n_{\rm H}$, and V denote the distance, electron density, hydrogen density, and volume, respectively.}
\end{table*}

\begin{figure}[hbt]
\figurenum{2}
\epsscale{0.99}
\plotone{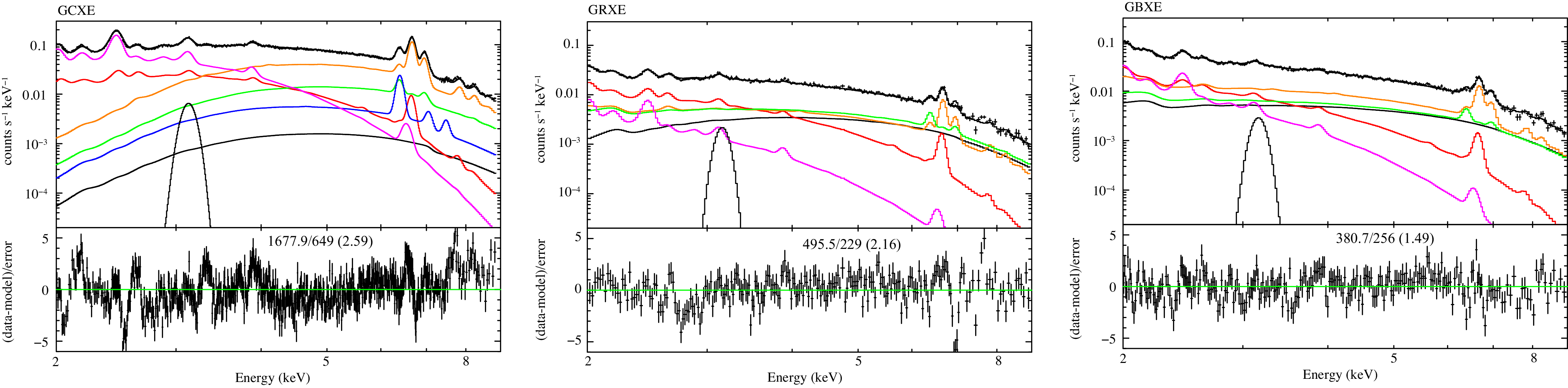}
\caption{
Top: GDXE (GCXE, GRXE. GBXE) spectra with the model~B, the combination of XAB (red), non-mCV (orange),  mCV
 (green), and aged CIE-SNR (magenta). 
Black power-law model exhibits the CXB.
Bottom: Residuals between spectrum and model, including a $\chi ^2$/d.o.f. (reduced~$\chi^2$) value.
}
\label{fig:modelB}
\end{figure}

\subsection {GCXE spectrum with aged CIE-SNRs and RP-SNRs (model C)} \label{sec:modelC}

In the fitting of the model B, significant residuals in GCXE are still observed, 
which is hardly reproduced by normal aged-SNR (CIE-SNRs). 
The SNR plasma in the sequence of $kT_{\rm i} < kT_{\rm e}$ is called as ionizing plasma (IP), hence SNR (IP-SNR). 
Normal evolution scenario of the IP-SNR follows the ionization sequence from low excited atoms (low $kT_{\rm i}$) to highly excited atom with high electron temperature (high $kT_{\rm e}$).
Then the IP-SNR finally reached to a collisional ionization equilibrium plasma (CIE) with $kT_{\rm i} = kT_{\rm e}$ 
(CIE-plasma). However, if there are some additional ionization, the CIE-plasma may be changed to 
ionization plasma with $kT_{\rm i} > kT_{\rm e}$. 
Then the IP plasma changes to recombining plasma (RP) or the SNR becomes an RP-SNR (e.g. \citealt{Ya21}).
 
Since residuals from the model B are likely due to highly ionized atoms in the recombining  plasma (RP), 
we added the aged RP-SNRs component in the model B (here we call as model~C), using the {\tt XSPEC} code of {\tt RNEI}. 
Since the RPs are generally accompanied with a neutral Fe~K$\alpha$ line (\citealt{No18, Ya21}), we further added a cold matter (CM) model, including a power-law ($\Gamma=2.1$) + neutral Fe-K$\alpha$ line with a free equivalent width of $EW_{6.4}$. The initial ionization temperature of the aged RP-SNRs were set to be $10~$keV. 
Then the ionization time-scale of $nt$ was searched in the range of $10^{10}$--$10^{13}$~cm$^{-3}$~s. 

\begin{figure}[hbt]
\figurenum{3}
\epsscale{0.5}
\plotone{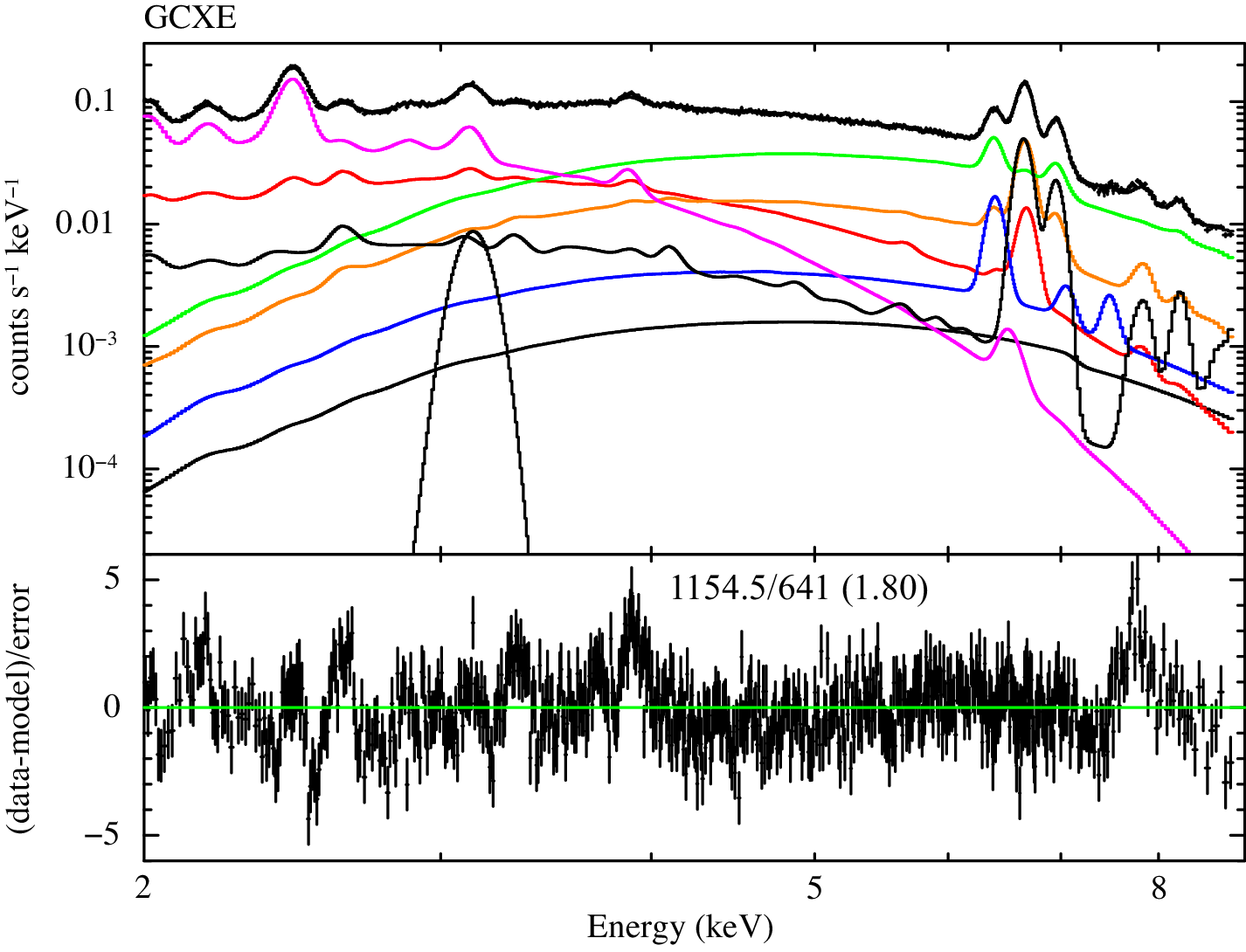}
\caption{
Top: GCXE spectrum with the model~C, the combination of XAB (red), non-mCV (orange),  mCV
 (green), aged CIE-SNR (magenta), aged RP-SNR (black), and CM (blue). 
Black power-law model exhibits the CXB.
Bottom: Residuals between spectrum and model, including a $\chi ^2$/d.o.f. (reduced~$\chi^2$) value.
}
\label{fig:modelC}
\end{figure}

The best-fit GDXE figures and their parameters are listed in Figure~\ref{fig:modelC} and in Table~\ref{tab:modelC}, respectively.
Notably, the reduced $\chi^2$ of 1.80 in the GCXE by model C is significantly improved from 2.59 (in model B). 
The best-fit k$T_{\rm e}$ and $z$ of the aged CIE-SNRs and RP-SNRs are consistent with  the high-quality {\it Suzaku} observations of aged-SNRs (e.g., \citealt{Ts09, Na10}). The best-fit aged RP-SNR yielded a $nt$ of $2.1(\pm0.3)\times 10^{10}$~cm$^{-3}$~s.

Notably the large abundance ratio of non-mCV found in model A and B (1.10/0.87) are disappeared in model C (0.88/0.87). 
The reason is  that the inclusion of the aged-SNR (either CIM or RP) cancels the apparent over abundance of $z$ 
in the non-mCV. 
Reasonable fractions of 6.4~keV, 6.7~keV, and 6.9~keV lines in the GCXE spectrum is originated from either CIE-SNRs or 
RP-SNRs, wherein the first one is obtained from the CM spectrum and the remaining ones are respectively derived from the aged RP-SNR.
The best-fit $EW_{6.4}$ in the CM is $\sim 1.1$~keV, and this value is more favorable for the low-energy cosmic-ray (LECR) ionization origin instead of the origin of hard X-ray (Figure~3 in \citealt{No15}; \citealt{Dogiel2011}). 
These are important roles of the aged-SNRs and CM for the GCXE. 
They improved the model C to be acceptable for the GRXE and GBXE or marginally acceptable for the GCXE.

\begin{table*} 
\caption{Best-fit parameters in model C.}
\label{tab:modelC}
\smallskip
\footnotesize
\begin{center}	
\begin{tabular}{llccccc}
\hline
&	&				&					& GC			& GR			& GB	\\
\hline
			& {\tt Absorption}	& ${N_{\rm H}} $ 	& $10^{22}$~cm$^{-2}$	& $5.93\pm 0.01$	& $2.12\pm 0.01$	& $0.75\pm0.01$	\\
\hline
{\bf XAB}		&{\tt Apec}	& $kT$		& keV				& $1.85\pm0.03$		& =GC			& =GC	\\
			&		& $Z^a$		& solar				& $0.32\pm0.01$	& 	=GC			& =GC	\\
			&		& $EM^b$	& $10^{-2}$			& $12.6\pm4.2$	& 	$0.32\pm0.07$		& $0.42\pm0.15$	\\
			& Fe~K$\alpha$	& $EW_{6.4}$	& eV				& $43$			& 	=GC			& =GC	\\
\hline
{\bf non-mCV}	&{\tt Apec}	& $kT$		& keV				& $5.0\pm0.1$	 	& =GC			& =GC	\\
			&		& $Z^a$		& solar			& $0.88\pm0.01$		& =GC			& =GC	\\
			& 		& $EM^b$	& $10^{-2}$			& $5.2\pm1.8$		& $0.17\pm0.05$		& $0.36\pm0.12$	\\
			& Fe~K$\alpha$	& $EW_{6.4}$	& eV				& $100$			& 	=GC			& =GC	\\
\hline
{\bf mCV}		&{\tt Apec}	& $kT$		& keV				& $20.0\pm0.1$		& 	=GC			& =GC	\\
			& 		& $Z^a$		& solar				& $0.41\pm0.01$	& 	=GC			& =GC	\\
			& 		& $EM^b$	& $10^{-2}$			& $9.7\pm1.6$		& 	$0.16\pm0.02$	& $0.28\pm0.05$	\\
			& Fe~K$\alpha$	& $EW_{6.4}$	& eV				& $160$			& 	=GC			& =GC	\\
\hline
{\bf CIE-SNR} 	&{\tt vApec}	& $kT$		& keV				& $0.83\pm0.03$		& 	=GC			& =GC	\\
			&		& $Z^a$		& solar			& $1.24\pm0.01$	 	& 	=GC			& =GC	\\
			&		& $EM^b$	& $10^{-2}$			& $53.3\pm7.5$		& 	$0.32\pm0.04$	& $0.87\pm0.12$	\\
\hline
{\bf RP-SNR} 	& {\tt vvRnei}& $nt$		& $10^{10}$~cm$^{-2}$~s	& $2.1\pm0.3$		&	---			& --- \\
			&		& $EM^b$	& $10^{-2}$			& $4.9\pm1.0$		& 	---			& ---	\\
\hline
{\bf CM} 		& Abs		& $N_{\rm H}$	& $10^{22}$~cm$^{-2}$	& $13.2\pm0.6$	& 	---				& ---	\\
			& PL		& $\Gamma$	& 				& $2.1\pm0.1$		& 	---			& ---	\\
			& Fe~K$\alpha$	& $EW_{6.4}$	& eV				& $1100\pm 200$		& 	---			& ---	\\
\hline			
\multicolumn{3}{c}{abundance ratio}                        	& ratio$^c$              &1.6                       	&1.33                      	&1.6  \\
\hline
\multicolumn{4}{c}{$\chi^2$/d.o.f.}			& 1154.5/641		& 343.6/229			& 408.1/256 \\
\hline
\end{tabular}
\end{center}
\tablenotetext{a}{Relative to abundances in table of \citealt{An89}.}
\tablenotetext{b}{Emission measure of plasma model, $\frac{10^{-14}}{4 \pi D^2} n_{\rm e} n_{\rm H} V$, where $D$, $n_{\rm e}$, $n_{\rm H}$, and V denote the distance, electron density, hydrogen density, and volume, respectively.}
\tablenotetext{c}{Abundance ratio of the parameter stated in Table 1.}
\end{table*}

\section{Summary and Discussion} \label{sec:sum-dis}
\subsection{Summary of GCXE, GRXE and GBXE based on XASs and SNRs}\label{sec:sum} 

In Section~3, we discussed the evolution history relating to the origin of GDXE based on the quantitative analysis. 
A quantitative examination for GDXE by composing of the XASs is rejected (model A, subsection 3.1). 
A unique trial of the comprehensive quantitative analysis was initially proposed by NK21.
In subsection 3.2, we follows the pioneering work of NK21 for all GDXEs adding CIE-SNRs in model A (called model B).
with the simultaneous fit of multi-{\tt Apec} model composed by XASs and aged CIE-SNRs.
The fit for the GDXE spectra are still rejected (see subsection~\ref{sec:modelB}).
In subsection 3.3, we confirm that  the combined  GDXE spectral fit  with  GCXE, GRXE, and GBXE 
by the assembly of real spectra of the XASs and aged RP-SNRs are improved significantly (see Table~\ref{tab:modelC}).  

The best fit by model C is marginally rejected for the GCXE, but are generally acceptable for the GRXE and GBXE.
Here, we should note that the model C is not a ``simple'' multicomponent plasma model, but its spectral parameters are highly constrained (not adjustable) by the real observed spectra of several XASs including aged-SNRs.
Furthermore, the best-fit parameters of each component were common among GCXE, GRXE, and GBXE. 
Thus, we conclude that the fits of the model C is not artificial chance, but it must be real fits for the GDXE 
by XASs and aged-SNRs (either in CIE or RP).   These indicate that the model C provides almost uniform and quantitative scenario to explain the entire GDXE spectra.

The RP-SNR in the GCXE originates from the ionization induced by the past big flare of Sgr A$^*$.  
Thus, the best-fit aged RP-SNR plasma of $nt \sim 2\times10^{10}$~cm$^{-3}$~s suggested a past Sgr~A$^*$ activity.
The best-fit $t(=nt/n)$ corresponds to $\sim 10^4$--$10^5$~years depending on the density of the aged RP-SNRs 
in the range of $n\sim 10^{-1}$--$10^{-2}$~cm$^{-3}$. 
The density $n$ was anti-correlated to the surface brightness of the aged RP-SNRs. The relevant aged RP-SNR 
should have low surface luminosity, or the $n$ should be $\lesssim 10^{-2}$~cm$^{-3}$. 
Then the ages of RP-SNR ($t$) are more than $\sim 10^5$~years, which is substantially larger than 
any epoch determined by the X-ray reflection nebula activity in Sgr~A, Sgr~B, and Sgr~C giant molecular clouds
(see figures in \citealt{Ry13} and \citealt{In09, Po15}).  
The big flare in the giant molecular clouds of Sgr~A--C should be appeared as a relics or precursor of 
several past flares of $\lesssim 10^{3}$ years (refer to \citealt{Ko86}), but no relics of the big flare are found in
Sgr~A--C. Note that the relics of the other possible precursors of Sgr~A$^*$ before $\gtrsim 10^5$~years would become unobservable in  Sgr~A--C, after the large recombination time scale of $t \sim 10^5$~years.

\subsection{Discussion on composed elements of GCXE, GRXE, and GBXE } \label{sec:dis}

\begin{figure}[htb]
\figurenum{4}
\epsscale{0.5}
\plotone{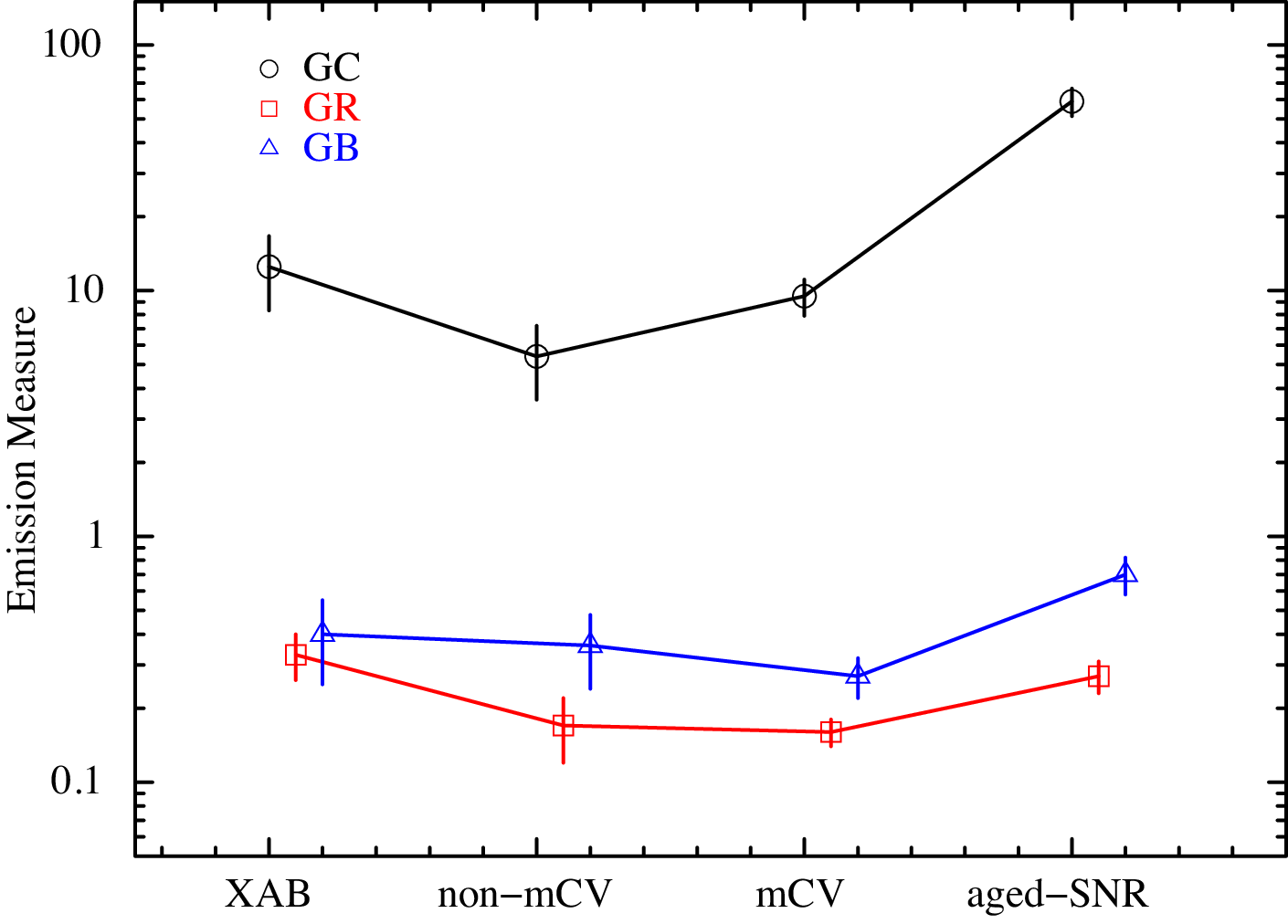}
\caption{
Best-fit $EM$ values of XAB, non-mCV, mCV, and aged SNR plotted in logarithmic scale, where those of aged SNRs are obtained from the sum of CIE and RP.
}
\label{fig:EM}
\end{figure}

To investigate the systematic difference of composition rate, we plot the $EM$ profile of non-mCV, mCV, XAB, and aged-SNRs for the GCXE, GRXE and GBXE.
The $EM$ profiles are plotted  in in logarithmic scale (Figure~\ref{fig:EM}).
As observed in Figure~\ref{fig:EM}, the $EM$ distribution of the XASs and aged-SNRs (CIE+RP)
in the GDXE is globally similar across the GCXE, GRXE, and GBXE.
However, upon detailed examination, we observe slight variations of the composition rates between the GCXE, GRXE, and GBXE.

These variations have been suggested by the scale height ($SH$) measurements of \cite{Uc13, Ya16}.  
The $SH$ near the GBXE regions includes two components of $\sim {1.5^\circ} $ and $\sim {0.5^\circ}$ \citep{Ya16}. 
The former is attributable to the aged-SNRs (either in CIE or RP) because of younger age compared to the  non-mCV and mCV, and hence smaller diffusion 
above the Galactic plane,  whereas the latter is mainly  due to non-mCVs and mCVs.
The latitude boundary $|b|$ in the GCXE, GRXE, and GBXE regions were ${0^\circ}$--$~{0.3^\circ}$, ${0^\circ}$--$~{1^\circ}$ and ${1.1^\circ}$--${2.8^\circ}$, respectively (refer to Table~\ref{tab:obsGDXE} in Appendix).
Therefore, depending on the mixing ratios of the non-mCVs and mCVs,  and aged-SNR between the GCXE, GRXE, and GBXE,
the $EM$ profiles among the GCXE, GRXE, and GBXE regions vary distinctly.    

The EM ratio of the aged-SNRs relative to the total (aged-SNRs $+$ XASs) gives useful information of the SN explosion rate.
\cite{Ko86} estimated the SN-rate of every 10 years on the extreme case that EM-ratio of aged-SNR is equal to 1 (or 100\%).
This paper determined more realistic EM-ratio to be  $\sim 33\%$. 
This new estimation shows that the SN rate is every $\sim 30$~years, which is more reasonable and reliable than the earlier estimation by \citet{Ko86}.
The best-fit model C parameters of the CM with the power-law index $\Gamma=2.1$ and $EW_{6.4}=1$~keV,  cannot precisely constrain the irradiation source, either through LECR protons or hard X-rays with ionization potential greater than 7.1~keV (e.g., \citealt{No15}).

In order to see that the reasonable fit by model C by multicomponent plasma with many adjustable parameters 
is not accidental but is real, we check one possibility of large systematic errors in long-time exposure observations. 
We re-analysis of the GCXE spectrum using $\sim 100$-ks exposure data from Obs.ID=505031010 
(Table~\ref{tab:obsGDXE} in Appendix). This exposure time is typical value in the previous $\chi^2$-test analyses. 
We apply the same fitting process on the results displayed in Figure~\ref{fig:modelC} as that for the original (long-exposure) GCXE spectrum. 
The best-fit $\chi^2$/d.o.f. is 445/416 (1.07),  which improves to 
a well acceptable level from the marginal level of 1.80 (see Table~\ref{tab:modelC}).  

The 100-ks spectrum improved the model C fit to be acceptable level for GCXE,  GRXE and GBXE.
One may claim that the error ranges are too small for the complicated spectra with many components in the model A, B and C fittings. The errors in Tables~1--4 are estimated 
by the {\tt error} command with 1-free parameter case. The spectra are high statistic data in Poisson distribution.  
In reality, the normalization of GCXE, GRXE and GBXE are also free parameters, hence we should estimate the errors by 3-free parameters of 
$kT_{\rm e}$, abundance $z$ and $N_{\rm H}$. 
Thus the estimated errors are given in Tables~2 and 3.  

As for the model C, the best-fit parameters are  found in Table~\ref{tab:modelC}. The error ranges of $kT_{\rm e}$, $z$ and $N_{\rm H}$ are similar with each other. Then the shape of error range in the three parameter
 ($kT_{\rm e}$, $z$, $N_{\rm H}$) space is roughly sphere, which means that the combined error is also small,  and is nearly the same as individual errors. 
This statement is also applicable to the 100-ks spectrum case.
The best-fit values and error ranges of each parameter are nearly the same between the long exposure spectrum and that of 100-ks exposure.

As we noted in section 3,  the large variation in the values and errors in the best fit  spectral parameters along the evolution starting from the model A to the models B and C shows large variation of $kT_{\rm e}$, while those of $z$ are almost constant. These facts are very important and suggestive for our further understandings of the true origin of GDXE by the high spectral resolution detector (X-ray micro-calorimeter) in the next Japanese satellite XRISM, launched on September 6, 2023 (UTC) \citep{Tashiro20}.

\begin{acknowledgments}
The authors deeply appreciate the {\it Suzaku} team providing high-quality data. This work is supported by JSPS KAKENHI Grant Numbers JP20H01742 and JP21K03615.
\end{acknowledgments}

\clearpage

\section*{Appendix: Observation Data}
The lists of {\it Suzaku} archive data for XASs (XAB, non-mCV, and mCV) and GDXE (GCXE, GRXE, and GBXE) used herein are summarized in Tables~\ref{tab:obsXAS} and \ref{tab:obsGDXE}, respectively.

\begin{table}[h]
\tablenum{5}
\begin{center}	
\caption{List of Observations of XAB, non-mCV and mCV.}
\label{tab:obsXAS}
\smallskip
\footnotesize
\begin{minipage}[t]{0.45\textwidth}
\centering
  \begin{tabular}{llcc}
      \hline
 Obs. ID  &Target & Distance$^a$ & Exposure$^b$  \\ 
 & & (pc) & (ks) \\
      \hline  \hline
\multicolumn{4}{l}{XAB} \\
\hline
 402095010 &GT MUS & 154 & 93.3 \\
 407038010 &II PEG & 40 & 111.5 \\
 402033010 &$\sigma$ GEM & 38 & 142.9 \\
 404008010 &UX ARIETIS & 52 & 87.8 \\
 401093010 &ALGOL & 28 & 102.2 \\
 402032010 &EV LAC & 5 & 34.2 \\
 401032010 &HR9024 & 130 & 58.8 \\
 405031010 &HD130693  & 30 & 21.3 \\
 401036010 &BETA LYR  & 295 & 20.3 \\
\hline \hline
\multicolumn{4}{l}{non-mCV} \\
\hline
 407047010 &BV CEN & 238  & 33.4 \\
 402046010 &BZ UMA & 228 & 29.8 \\
 407044010 &EK TRA & 180 & 77.9 \\
 403039010 &FL PSC & 145 & 33.3 \\
 408041010 &FS AUR & 578 & 62.2 \\
 403041010 &KT PER & 145 & 29.2 \\
 402045010 &SS AUR & 201 & 19.5 \\
 400007010 &SS CYG & 166 & 56.1 \\
 407034010 &U GEM & 100 & 119.1 \\
 408029010 &V1159 ORI & 357 & 200.6 \\
 401041010 &V893 SCO & 135 & 18.5 \\
 406009010 &VW HYI & 64  & 70.1 \\
 402043010 &VY AQR & 97 & 25.4 \\
 404022010 &Z CAM & 163 & 37.7 \\
\hline 
    \end{tabular}
\end{minipage}
\begin{minipage}[t]{0.45\textwidth}
\centering
  \begin{tabular}{llcc}
      \hline
 Obs. ID  &Target & Distance$^a$ & Exposure$^b$  \\ 
  & & (pc) & (ks) \\
\hline \hline
\multicolumn{4}{l}{mCV} \\
\hline 
  400016020 &CH CYG  & 268 & 33.3 \\
 406033010 &RS OPH & 1400 & 69.4 \\
 402040010 &RT CRU & 1800 & 50.9 \\
 401055010 &SS73 17 & 500 & 19.2 \\
 401043010 &T CRB & 1060 & 46.3 \\
 905001010 &V407 CYG & 4000 & 42.2 \\
 404033010 &AO PSC & 330 & 39.7 \\
 404029010 &BG CMI & 890 & 47.1 \\
 402001010 &EX HYA  & 65 & 100.5 \\
 404032010 &FO AQR & 450 & 46.1 \\
 403081010 &GK PER & 420 & 30.4 \\
 403028010 &IGR J17195$-$4100 & 640 & 31.6 \\
 403026010 &IGR J17303$-$0601 & 2290 & 33.0 \\
 403004010 &MU CAM & 953 & 50.3 \\
 401037010 &NY LUP & 680 & 99.5 \\
 404030010 &PQ GEM & 510 & 46.7 \\
 403023010 &TV COL & 368 & 35.8 \\
 404031010 &TX COL & 909 & 59.8 \\
 402002010 &V1223 SGR & 527 & 60.7 \\
 401038010 &1RXS J213344.1$+$51072 & 1420 & 81.9 \\
 403021010 &V2400 Oph & 280 & 110.5 \\
 403025010 &V709 CAS & 230 & 35.9 \\
  500015010 &XY ARI & 270 & 102.9 \\
 403007010 &AM HERCULES & 91 & 108.5 \\
 403027010 &V1432 Apl & 230 & 32.5 \\
 408030010 &SWIFT J2319.4$+$2619 & 520 & 41.3 \\
      \hline
    \end{tabular}
\end{minipage}
\end{center}
\tablenotetext{a}{Distance to the source referred to Simbad \citep{We00}.  }
\tablenotetext{b}{Effective exposure time. }
\end{table}

\begin{table}[h]
\tablenum{6}
\begin{center}
\caption{List of the GDXE observations.}
\label{tab:obsGDXE}
\smallskip
\scriptsize
\begin{minipage}[t]{0.45\textwidth}
\centering
  \begin{tabular}{lcc}
      \hline
Obs. ID  & Position ($l$, $b$) 	& Exposure (ks)$^a$  \\  
      \hline \hline
\multicolumn{3}{l}{GCXE} \\
      \hline
100027010 & ({0.039$^\circ$}, {-0.088$^\circ$}) & 44.8 \\
100027020 & ({359.736$^\circ$}, {-0.059$^\circ$}) & 42.8 \\
100037010 & ({359.737$^\circ$}, {-0.059$^\circ$}) & 43.7 \\
100037040 & ({0.038$^\circ$}, {-0.087$^\circ$}) & 43.0 \\
100048010 & ({0.037$^\circ$}, {-0.086$^\circ$}) & 63.0 \\
102013010 & ({0.037$^\circ$}, {-0.087$^\circ$}) & 51.4 \\
408017090 & ({359.944$^\circ$}, {-0.045$^\circ$}) & 22.2 \\
409011010  & ({359.943$^\circ$}, {-0.049$^\circ$}) & 20.2 \\
409011020 & ({359.946$^\circ$}, {-0.050$^\circ$}) & 17.3 \\
500005010 & ({0.445$^\circ$}, {-0.099$^\circ$}) & 88.4 \\
500018010 & ({359.451$^\circ$}, {-0.077$^\circ$}) & 106.9 \\
501008010 & ({359.825$^\circ$}, {-0.205$^\circ$}) & 129.6 \\
501009010 & ({359.906$^\circ$}, {0.164$^\circ$}) & 51.2 \\
502022010 & ({0.209$^\circ$}, {-0.284$^\circ$}) & 134.8 \\
503007010 & ({0.313$^\circ$}, {0.151$^\circ$}) & 52.2 \\
503072010 & ({359.588$^\circ$}, {0.188$^\circ$}) & 140.6 \\
505031010 & ({359.523$^\circ$}, {-0.132$^\circ$}) & 100.0 \\
508019010 & ({359.413$^\circ$}, {-0.109$^\circ$}) & 104.2 \\
508064010 & ({0.038$^\circ$}, {-0.090$^\circ$}) & 50.5 \\
\hline 
\multicolumn{2}{l}{total} & 1306.8 \\
\hline \\
      \hline \hline
\multicolumn{3}{l}{GBXE} \\
      \hline
502059010 & ( {0.000$^\circ$}, {-2.002$^\circ$} ) & 136.8 \\
503099010 & ( {359.780$^\circ$}, {1.134$^\circ$} ) & 29.7 \\
503100010 & ( {359.310$^\circ$}, {1.134$^\circ$} ) & 25.7 \\
504089010 & ( {359.951$^\circ$}, {-1.202$^\circ$} ) & 55.3 \\
504090010 & ( {358.506$^\circ$}, {-1.185$^\circ$} ) & 41.3 \\
504091010 & ( {358.500$^\circ$}, {-1.603$^\circ$} ) & 51.3 \\
504093010 & ( {358.500$^\circ$}, {-2.802$^\circ$} ) & 53.2 \\
505078010 & ( {359.956$^\circ$}, {-1.596$^\circ$} ) & 51.3 \\
505079010 & ( {359.955$^\circ$}, {-2.793$^\circ$} ) & 50.2 \\
507028010 & ( {0.298$^\circ$}, {-1.546$^\circ$} ) & 51.7 \\
507029010 & ( {0.220$^\circ$}, {-1.870$^\circ$} ) & 52.5 \\
507030010 & ( {0.567$^\circ$}, {-1.469$^\circ$} ) & 51.7 \\
507031010 & ( {0.459$^\circ$}, {-1.770$^\circ$} ) & 52.9 \\
509080010 & ( {359.836$^\circ$}, {1.452$^\circ$} ) & 88.2 \\
\hline 
\multicolumn{2}{l}{total} & 791.8 \\
\hline \\
\\
\\
\\
\\
\\
\\
\\
\\
\\
\\
\\
\\
\\
\\
\\
\\
    \end{tabular}
\end{minipage}
\begin{minipage}[t]{0.45\textwidth}
\centering
  \begin{tabular}{lcc}
      \hline
Obs. ID  & Position ($l$, $b$) 	& Exposure (ks)$^a$  \\ 
      \hline \hline
\multicolumn{3}{l}{GRXE} \\
      \hline
100026020 & ( {347.632$^\circ$}, {0.708$^\circ$} ) & 34.9 \\
100026030 & ( {345.806$^\circ$}, {-0.541$^\circ$} ) & 37.5 \\
100028010 & ( {332.404$^\circ$}, {-0.150$^\circ$} ) & 41.4 \\
100028020 & ( {332.003$^\circ$}, {-0.150$^\circ$} ) & 19.3 \\
100028030 & ( {332.704$^\circ$}, {-0.150$^\circ$} ) & 21.9 \\
401026010 & ( {25.266$^\circ$}, {-0.103$^\circ$} ) & 42.2 \\
401054010 & ( {341.319$^\circ$}, {0.578$^\circ$} ) & 21.2 \\
401056010 & ( {333.491$^\circ$}, {0.300$^\circ$} ) & 39.1 \\
401101010 & ( {12.870$^\circ$}, {0.012$^\circ$} ) & 63.8 \\
404056010 & ( {338.002$^\circ$}, {0.079$^\circ$} ) & 50.6 \\
404081010 & ( {29.766$^\circ$}, {-0.205$^\circ$} ) & 104.3 \\
405027010 & ( {340.439$^\circ$}, {-0.178$^\circ$} ) & 21.0 \\
406069010 & ( {10.003$^\circ$}, {-0.234$^\circ$} ) & 70.6 \\
406078010 & ( {340.168$^\circ$}, {-0.117$^\circ$} ) & 149.8 \\
407018010 & ( {333.606$^\circ$}, {-0.205$^\circ$} ) & 40.5 \\
407020010 & ( {333.723$^\circ$}, {0.218$^\circ$} ) & 44.3 \\
407091010 & ( {333.891$^\circ$}, {0.409$^\circ$} ) & 29.3 \\
500009010 & ( {28.458$^\circ$}, {-0.212$^\circ$} ) & 93.3 \\
500009020 & ( {28.457$^\circ$}, {-0.211$^\circ$} ) & 98.9 \\
501042010 & ( {331.575$^\circ$}, {-0.528$^\circ$} ) & 40.2 \\
501043010 & ( {330.400$^\circ$}, {-0.377$^\circ$} ) & 43.6 \\
501044010 & ( {17.871$^\circ$}, {-0.703$^\circ$} ) & 50.3 \\
501045010 & ( {18.437$^\circ$}, {-0.844$^\circ$} ) & 52.2 \\
501105010 & ( {348.797$^\circ$}, {-0.535$^\circ$} ) & 20.7 \\
502001010 & ( {11.949$^\circ$}, {-0.090$^\circ$} ) & 53.8 \\
502049010 & ( {344.262$^\circ$}, {-0.220$^\circ$} ) & 215.7 \\
502053010 & ( {15.820$^\circ$}, {-0.848$^\circ$} ) & 71.5 \\
503028010 & ( {17.608$^\circ$}, {-0.840$^\circ$} ) & 57.2 \\
503029010 & ( {17.732$^\circ$}, {-0.440$^\circ$} ) & 57.2 \\
503030010 & ( {17.469$^\circ$}, {-0.578$^\circ$} ) & 55.5 \\
503073010 & ( {331.299$^\circ$}, {-0.761$^\circ$} ) & 53.7 \\
503074010 & ( {331.466$^\circ$}, {-0.636$^\circ$} ) & 52.7 \\
503078010 & ( {11.029$^\circ$}, {0.072$^\circ$} ) & 51.6 \\
503079010 & ( {10.844$^\circ$}, {0.043$^\circ$} ) & 44.3 \\
503086010 & ( {18.004$^\circ$}, {-0.695$^\circ$} ) & 52.1 \\
503108010 & ( {348.921$^\circ$}, {-0.454$^\circ$} ) & 23.6 \\
504052010 & ( {26.447$^\circ$}, {0.132$^\circ$} ) & 41.0 \\
504077010 & ( {11.609$^\circ$}, {-0.253$^\circ$} ) & 51.9 \\
504078010 & ( {11.327$^\circ$}, {-0.060$^\circ$} ) & 52.5 \\
504079010 & ( {10.715$^\circ$}, {0.332$^\circ$} ) & 51.0 \\
504099010 & ( {25.502$^\circ$}, {0.007$^\circ$} ) & 52.7 \\
505025010 & ( {22.000$^\circ$}, {0.004$^\circ$} ) & 50.5 \\
505026010 & ( {23.492$^\circ$}, {0.039$^\circ$} ) & 49.0 \\
505052010 & ( {340.771$^\circ$}, {-1.013$^\circ$} ) & 49.6 \\
505076010 & ( {347.853$^\circ$}, {-0.227$^\circ$} ) & 32.6 \\
505088010 & ( {26.312$^\circ$}, {-0.001$^\circ$} ) & 49.6 \\
505089010 & ( {26.397$^\circ$}, {-0.311$^\circ$} ) & 50.0 \\
505090010 & ( {26.705$^\circ$}, {-0.152$^\circ$} ) & 49.6 \\
505091010 & ( {27.128$^\circ$}, {-0.280$^\circ$} ) & 51.2 \\
506021010 & ( {23.299$^\circ$}, {0.310$^\circ$} ) & 40.3 \\
506051010 & ( {18.785$^\circ$}, {0.397$^\circ$} ) & 52.1 \\
507044010 & ( {19.565$^\circ$}, {0.007$^\circ$} ) & 171.8 \\
904006010 & ( {23.403$^\circ$}, {0.039$^\circ$} ) & 42.4 \\
\hline
\multicolumn{2}{l}{total} & 2957.6 \\
      \hline \\

    \end{tabular}
\end{minipage}
\end{center}
\tablenotetext{a}{Effective exposure time. }
\end{table}

\clearpage

\bibliography{Draft_bib_2022-1228}{}
\bibliographystyle{aasjournal}

\end{document}